\documentclass[3p,twocolumn,sort&compress,longbibliography,fleqn]{elsarticle}
\usepackage{amsmath, bm}
\usepackage{amssymb}
\usepackage{braket}
\usepackage{listings}
\usepackage{cprotect}
\usepackage{lmodern}
\usepackage{xspace}
\usepackage{ulem}
\usepackage{color}
\usepackage{physics}
\usepackage{cprotect}

\journal{Computer Physics Communications}

\newcommand{\software}{\texttt{wan2respack}\xspace}
\def\vec#1{\boldsymbol #1}

\usepackage{hyperref}
\hypersetup{colorlinks=true,breaklinks,linkcolor=blue,urlcolor=blue,citecolor=blue}
\usepackage{listings}

\long\def\beginmypgfpdfnamed#1#2\endmypgfpdfnamed{\includegraphics{#1}}

\graphicspath{{./pyfig/}}

\begin{document}

\begin{frontmatter}

\title{Interface tool from Wannier90 to RESPACK: \software}

\author[tohoku]{Kensuke~Kurita}
\author[issp,baqis]{Takahiro~Misawa}
\author[issp]{Kazuyoshi~Yoshimi}
\author[issp]{Kota~Ido}
\author[tohoku]{Takashi~Koretsune}

\address[tohoku]{Department of Physics, Tohoku University, Sendai 980-8578, Japan}
\address[issp]{The Institute for Solid State Physics, The University of Tokyo, Chiba 277-8581, Japan}
\address[baqis]{Beijing Academy of Quantum Information Sciences, Haidian District, Beijing 100193, China}

\begin{abstract}
We develop the interface tool \verb|wan2respack|, which connects \verb|RESPACK| (software that derives the low-energy effective Hamiltonians of solids) with \verb|Wannier90| (software that constructs Wannier functions). \verb|wan2respack| converts the Wannier functions obtained by \verb|Wannier90| into those used in \verb|RESPACK|, which is then used to derive the low-energy effective Hamiltonians of solids. In this paper, we explain the basic usage of \verb|wan2respack| and show its application to 
standard compounds of correlated materials, namely, the correlated metal SrVO$_3$ and the high-$T_{c}$ superconductor La$_2$CuO$_4$. Furthermore, we compare the low-energy effective Hamiltonians of these compounds using Wannier functions obtained by \verb|Wannier90|  and those obtained by \verb|RESPACK|. 
We confirm that both types of Wannier functions give the same Hamiltonians. This benchmark comparison demonstrates that \verb|wan2respack| correctly converts Wannier functions in the \verb|Wannier90| format into those in the \verb|RESPACK| format.
\end{abstract}

\begin{keyword}
Wannier functions, $ab$ $initio$ downfolding, constrained random phase approximation,
strongly correlated electron systems
\end{keyword}

\end{frontmatter}
\noindent
{\bf PROGRAM SUMMARY}

\begin{small}
\noindent
{\em Program title:}  \software\\
{\em Licensing provisions:} GNU General Public License version 3\\
{\em Programming language:} \verb*#Fortran# and  \verb*#python3#\\
{\em Computer:} PC, cluster machine\\ 
{\em Operating system:} Unix-like system, tested on Linux and macOS\\ 
{\em Keywords:} Wannier functions, $ab$ $initio$ downfolding, constrained random phase approximation, strongly correlated electron systems.\\
{\em External routines/libraries:} \verb|Quantum ESPRESSO| (version 6.6), \verb|Wannier90| (version 3.0.0),  \verb|RESPACK| (version 20200113), \verb|tomli|. \\
{\em Nature of problem:} 
Using \verb|RESPACK|, one can derive low-energy effective Hamiltonians of solids from maximally localized Wannier functions. However, due to the differences in the representation of Wannier functions, the Wannier functions obtained by \verb|Wannier90| cannot be directly used in \verb|RESPACK|.
\\
{\em Solution method:}
\software converts the
Wannier functions in the \verb|Wannier90| format into those in the \verb|RESPACK| format. 
Using the converted  Wannier functions, one can derive the low-energy effective Hamiltonians using \verb|RESPACK|.
\\
\end{small}

\section{Introduction}
Wannier functions, constructed through unitary transformations of Bloch functions, offer simple, convenient representations of the electronic structures of solids~\cite{Marzari_RMP2012}. After the development of an efficient method of obtaining unique sets of maximally localized Wannier functions (MLWFs)~\cite{Marzari_PRB1997,Souza_PRB2001}, MLWFs have been widely used to analyze the electronic structures of solids. For example, MLWFs play an essential role in calculating electronic polarizations~\cite{King-Smith_PRB1993,Resta_RMP1994} and orbital magnetizations~\cite{Thonhauser_PRL2005,Xiao_PRL2005}. 
MLWFs are also used to construct tight-binding models~\cite{Yates_PRB2007}, which reproduce low-energy bands near the Fermi level.

MLWFs are also vital for deriving the $ab$ $initio$ low-energy effective Hamiltonians of solids ~\cite{Aryasetiawan_PRB2004,Imada_JPSJ2010}. Based on MLWFs, the screened interactions in low-energy effective Hamiltonians are evaluated through constrained random phase approximation (cRPA). The derivation of low-energy effective Hamiltonians is often called $ab$ $initio$ downfolding. Correlation effects beyond the conventional density functional theory (DFT) \cite{PhysRev.136.B864, PhysRev.140.A1133} can be considered by solving low-energy effective Hamiltonians using accurate solvers. In the last decade, this method has been applied to a wide range of strongly correlated compounds~\cite{Nakamura_JPSJ2008,Nakamura_PRB2009,Nakamura_JPSJ2009,Miyake_JPSJ2010,Misawa_JPSJ2011,Nohara_JPSJ2011,Wehling_PRL2011,Saioglu_PRL2012,Nakamura_PRB2012,Nomura_PRB2012,Vaugier_PRB2012,Arita_PRL2012,Shinaoka_JPSJ2012,Nilson_PRB2013,Hansmann_JPS2013,Yamaji_PRL2014,Okamoto_PRB2014,Amadon_PRB2014,Kim_PRB2016,Seth_PRL2017,Hirayama_PRB2018,Moree_PRB2018,Tadano_PRB2019,Hirayama_PRB2019,Nomura_PRB2019,Hirayama_PRB2020}.

The open-source software package \verb|Wannier90|~\cite{Mostofi_CPC2008,Mostofi_CPC2014,Pizzi_JPCM2020,wan90_HP} was developed to construct MLWFs from the results of $ab$ $initio$ band calculations. The MLWFs obtained by \verb|Wannier90| can be used to calculate many important properties of solids, such as Berry phases and anomalous Hall conductivity~\cite{Wang_PRB2006}, electrical conductivity~\cite{Pizzi_CPC2014}, and spin Hall conductivity~\cite{Qiao_PRB2018}. Because of this versatility, many $ab$ $initio$ calculation software packages, such as \verb|Quantum ESPRESSO|~\cite{QE,QE_HP}, \verb|VASP|\cite{KRESSE199615,PhysRevB.54.11169}, \verb|WIEN2k|\cite{doi:10.1063/1.5143061}, \verb|ABINIT|\cite{GONZE2020107042}, and \verb|OpenMX|\cite{OpenMX_HP,Ozaki_PRB2003,Ozaki_PRB2004}, have an interface to \verb|Wannier90|. 

Nakamura {\it et al}. recently developed the open-source software package \verb|RESPACK|~\cite{Nakamura_CPC2021,respack_HP}, which implements $ab$ $initio$ downfolding. After its release, \verb|RESPACK| has been applied to a wide range of correlated materials~\cite{Huang_PRR2020,Misawa_PRR2020,Masuda_PRB2021,Charlebois_PRB2021,Ido_arXiv2021,Yoshimi_PRR2021,Morishita_APE2021,Steinhoff_PRB2021,Kanno_2021JPCL,Ido_npj2022,Moree_PRB2022,Huebsch_2022JPCM,Li_2022EL,Ohki_arXiv2022,Yoshimi_arXiv2022}.
\verb|RESPACK| also provides functions (\verb|RESPACK|-Wannier) that construct MLWFs using the $ab$ $initio$ band calculation results obtained by \verb|Quantum ESPRESSO| or \verb|xTAPP|~\cite{TAPP,xTAPP}. The \verb|RESPACK|-Wannier code was developed independently of \verb|Wannier90|. Therefore, their representation of Wannier functions has 
technical differences, such as the choice of $k$-point mesh (e.g., based on irreducible or reducible representation), although the constructed Wannier functions themselves are equivalent. Due to these differences, \verb|RESPACK| cannot be directly connected with \verb|Wannier90|.

In this paper, we introduce the interface tool \verb|wan2respack|~\cite{wan2respack}, which converts Wannier functions in the \verb|Wannier90| format into those in the \verb|RESPACK| format. After the conversion of Wannier functions, the screened Coulomb and exchange interactions can be evaluated using \verb|RESPACK|. The rest of this paper is organized as follows. In Section 2, we give an overview of the formats of the MLWFs implemented in \verb|Wannier90| and \verb|RESPACK| and explain how to convert them using \verb|wan2respack|. We also explain how to derive low-energy effective Hamiltonians based on the converted MLWFs. In Sections 3 and 4, we explain the installation and usage, respectively, of \verb|wan2respack|. In Section 5, we show the application of \verb|wan2respack| to the correlated materials SrVO$_3$ and La$_2$CuO$_4$. Section 6 summarizes this paper.

\section{Overview}
This section provides an overview of the formats of the Wannier functions implemented in \verb|Wannier90| and \verb|RESPACK|. We also explain how \verb|wan2respack| converts Wannier functions in the \verb|Wannier90| format into those in the \verb|RESPACK| format. Finally, we discuss how to derive the low-energy effective Hamiltonians of solids based on the converted MLWFs.

We obtain MLWFs through unitary transformations of Bloch functions. The Wannierization procedure consists of the following optimization of two quantities:
\begin{enumerate}
\item Projection [$U({\bm k})^{\rm opt}_{mn}$ in Eq.~(\ref{eq:wannier_w90})] of the Bloch functions within the given energy window to minimize the gage-invariant Wannier spread
\item Unitary transformation [$U_{im}({\bm k})$ in Eq.~(\ref{eq:wannier_w90_opt})] of the optimal Bloch wave functions ($\psi^{\rm opt}({\bm k})$ in Eq.~(\ref{eq:wannier_w90_opt})) obtained by the first step
\end{enumerate}
According to the Wannierization procedure, the $i$th Wannier function around site $\bm R$ ($\bm R$ is often called the Wannier center), $w_{i \bm R}(\bm r) = w_{i \bm 0}(\bm r - \bm R)$, is defined as
\begin{align}
w_{i\bm R}({\bm r}) = \frac{1}{\sqrt{N_{\bm k}}} \sum_{{\bm k}} \sum_{m} U_{im}({\bm k}) \psi^{\rm opt}_{m{\bm k}}({\bm r}) e^{-i \bm k \cdot \bm R},
\label{eq:wannier_w90}
\end{align}
where $N_{\bm k}$ is the number of $\bm k$-point meshes and $U_{im}({\bm k})$ is the unitary matrix, which transforms the $m$th optimized Bloch wave function $\psi_{m \bm k}^{\rm opt}({\bm r})$ into the $i$th Wannier function. Here, $\psi_{m \bm k}^{\rm opt}({\bm r})$ is given by
\begin{align}
\psi^{\rm opt}_{m{\bm k}}({\bm r}) = \sum_{n} U_{mn}^{\rm opt}({\bm k}) \psi_{n{\bm k}}(\bm r), 
\label{eq:wannier_w90_opt}
\end{align}
where $U_{mn}^{\rm opt}({\bm k})$ is the projection matrix. The Bloch wave function $\psi_{n \bm k}({\bm r})$ is defined as
\begin{align}
 \psi_{n \bm k}({\bm r}) =\frac{1}{\sqrt{N_k}} \frac{1}{\sqrt{\Omega}} \sum_{\bm G} C_{\bm Gn}(\bm k) e^{i(\bm k+ \bm G)\cdot \bm r},
\end{align}
where $\Omega$ is the volume of the calculation unit cell and $C_{\bm Gn}(\bm k)$ is the expansion coefficient of the plane wave. 

Here, we explain the difference between the \verb|Wannier90| and \verb|RESPACK| formats of Wannier functions. By substituting the definition of the Bloch functions into Eq. (\ref{eq:wannier_w90}) for $\bm{R}=\bm{0}$, we obtain
\begin{align}
w_{i0}(\bm r) = 
\frac{1}{N_k} \sum_{\bm k} \sum_{\bm G} \tilde{C}_{\bm G i}(\bm k)\frac{1}{\sqrt{\Omega}}e^{i(\bm k+\bm G)\cdot \bm r}.
\label{eq:wan_respack}
\end{align}
The relation between $\tilde{C}_{\bm G i}(\bm k)$ and $C_{\bm G n}$ is 
\begin{align}
\tilde{C}_{\bm G i}(\bm k) = \sum_{m,n}U_{im}({\bm k})U_{mn}^{\rm opt}({\bm k})C_{\bm G n}(\bm k).
\end{align}
The Wannier functions in \verb|RESPACK| are defined in Eq.~(\ref{eq:wan_respack}) and used in the calculations of the screened Coulomb interactions. \verb|RESPACK| reads wave functions in the irreducible Brillouin zone (IBZ), $C_{\bm G n}(\bm k)$, where $\bm k \in {\rm IBZ}$, to reduce the computational cost, particularly that of the dielectric function. Since crystal symmetry is not taken into account in the standard Wannierization procedure\cite{note_sym}, $U_{im}(\bm k)$, $U_{mn}^{\rm opt}(\bm k)$, and $\tilde{C}_{\bm G i}(\bm k)$ are calculated on the full $\bm k$-point mesh by expanding $C_{\bm G n}(\bm k)$ in the IBZ to the full $\bm k$-point mesh. When using \verb|wan2respack|, we first calculate wave functions in the IBZ, $C_{\bm G n}(\bm k)$, to compute the dielectric function. Then, we separately calculate wave functions on the full $\bm k$-point mesh and generate Wannier functions using \verb|Quantum ESPRESSO| and \verb|Wannier90|. \verb|wan2respack| computes $\tilde{C}_{\bm G i}(\bm k)$ based on these outputs. Here, we have to use the same full $\bm k$-point mesh used in \verb|RESPACK|.

After the Wannier functions are obtained, the following $ab$ $initio$ Hamiltonians are derived:
\begin{align}
&H=\sum_{\sigma,\vec{R}\vec{R}^{\prime},\alpha\beta}
t_{\alpha\vec{R},\beta\vec{R}^{\prime}}
c_{\alpha\vec{R}\sigma}^{\dagger}c_{\beta\vec{R}^{\prime}\sigma} \notag \\
&+\frac{1}{2}\sum_{\sigma\rho,\vec{R}\vec{R}^{\prime},\alpha\beta}
\Big[U_{\alpha\vec{R},\beta\vec{R}^{\prime}}
c_{\alpha\vec{R}\sigma}^{\dagger}c_{\beta\vec{R}^{\prime}\rho}^{\dagger}
c_{\beta\vec{R}^{\prime}\rho}c_{\alpha\vec{R}\sigma} \notag\\
&+J_{\alpha\vec{R},\beta\vec{R}^{\prime}}
(c_{\alpha\vec{R}\sigma}^{\dagger}c_{\beta\vec{R}^{\prime}\rho}^{\dagger}
c_{\beta\vec{R}\rho}c_{\alpha\vec{R}^{\prime}\sigma} \notag\\
&+c_{\alpha\vec{R}\sigma}^{\dagger}c_{\alpha\vec{R}\rho}^{\dagger}
c_{\beta\vec{R}^{\prime}\rho}c_{\beta\vec{R}^{\prime}\sigma} 
)\Big],
\end{align}
where $c_{\alpha\vec{R}\sigma}^{\dagger}$ ($c_{\alpha\vec{R}\sigma}$) is a creation (annihilation) operator of the $\alpha$th Wannier orbital on $\vec{R}$ with the spin $\sigma$. The transfer integral and the screened Coulomb and exchange interactions are defined as 
\begin{align}
t_{\alpha\vec{R},\beta\vec{R}^{\prime}}
&=\int_{V}d\vec{r}~w_{\alpha\vec{R}}^{*}(\vec{r})H_{0}(\vec{r})w_{\beta\vec{R}^{\prime}}(\vec{r}) \label{eq:t}, \\
U_{\alpha\vec{R},\beta\vec{R}^{\prime}}&=
\int_{V}d\vec{r}d\vec{r}^{\prime}
w_{\alpha\vec{R}}^{*}(\vec{r})w_{\alpha\vec{R}}(\vec{r}) \notag \\
&\times W(\vec{r},\vec{r}^{\prime})
w_{\beta\vec{R}^{\prime}}^{*}(\vec{r}^{\prime})w_{\beta\vec{R}^{\prime}}(\vec{r}^{\prime})
\label{eq:U}, \\
J_{\alpha\vec{R},\beta\vec{R}^{\prime}}
&=\int_{V}d\vec{r}d\vec{r}^{\prime}
w_{\alpha\vec{R}}^{*}(\vec{r})w_{\beta\vec{R}^{\prime}}(\vec{r}) \notag \\
&\times W(\vec{r},\vec{r}^{\prime})
w_{\beta\vec{R}^{\prime}}^{*}(\vec{r}^{\prime})w_{\alpha\vec{R}}(\vec{r}^{\prime}).
\label{eq:J}
\end{align}
Here, $H_0(\bm r)$ and $W(\bm r, \bm r')$ represent the one-body part of the Hamiltonian (Kohn--Sham Hamiltonian) and the static screened Coulomb interaction obtained through cRPA, respectively. The integrals are obtained over the crystal volume $V$. \verb|RESPACK| calculates these quantities based on the MLWFs.

\section{Installation of \software}
\software can be downloaded from the following GitHub repository:
\begin{verbatim}
https://github.com/respack-dev/wan2respack
\end{verbatim}
Users can compile \verb|wan2respack| using \verb|CMake|. The typical installation procedure of \verb|wan2respack| is as follows:
\begin{verbatim}
$ cd $PATH_to_wan2respack
$ mkdir build
$ cd build
$ cmake ../ -DCONFIG=$Type_of_Configure 
-DCMAKE_INSTALL_PREFIX=$PATH_to_Install
$ make
$ make install 
\end{verbatim}
\verb|$PATH_to_wan2respack| is the path to the \software directory and \verb|$PATH_to_Install| is the path to the installation directory. By replacing \verb|$Type_of_Configure| with the name of \verb|CMake| configuration files, users can specify their desired compilers. 

All binary files and Python scripts are installed to \verb|$PATH_to_Install/bin|. 
The roles of the scripts in \verb|bin| are as follows:
\begin{itemize}
    \item \verb|init.py|: Common functions are defined in this Python module.
    \item \verb|qe2respack.py|: This Python script is used to generate the input files of \verb|RESPACK| from the output files obtained by \verb|Quantum ESPRESSO| band calculations.\cprotect\footnote{This script is originally distributed under GNU GPL version 3 by the open-source software \verb|RESPACK| version 20200113.}
    
    \item \verb|wan2respack.py|: This is the main Python script. The configuration file name should be written in the \verb|toml| format~\cite{toml}. This script calls \verb|wan2respack_pre.py| for preprocessing and \verb|wan2respack_core.py| for core processing. These scripts are described as follows:
\begin{itemize}
    \item \verb|wan2respack_pre.py|: This Python script is used to save the \verb|Quantum ESPRESSO| results and export ${\bm k}$ points with \verb|gen_mk.x| and \verb|qe2respack.py|.
    \begin{itemize}
    \item \verb|gen_mk.x|: This execution file, written in \verb|Fortran90|, for calculating the ${\bm k}$-point mesh for \verb|Wannier90|. 
    \end{itemize}
    
    \item \verb|wan2respack_core.py|: This Python script is used to prepare files about Wannier functions in the \verb|RESPACK| format using \verb|gen_wan.x| and \verb|qe2respack.py|.
    \begin{itemize}
            \item \verb|gen_wan.x|: This execution file, written in \verb|Fortran90|, converts the Wannier function information in the \verb|Wannier90| format into that in the \verb|RESPACK| format.
    \end{itemize}
    \end{itemize}

\end{itemize}

\section{Usage of \software}
\begin{figure}[tb]
    \centering
    \includegraphics[width=\columnwidth]{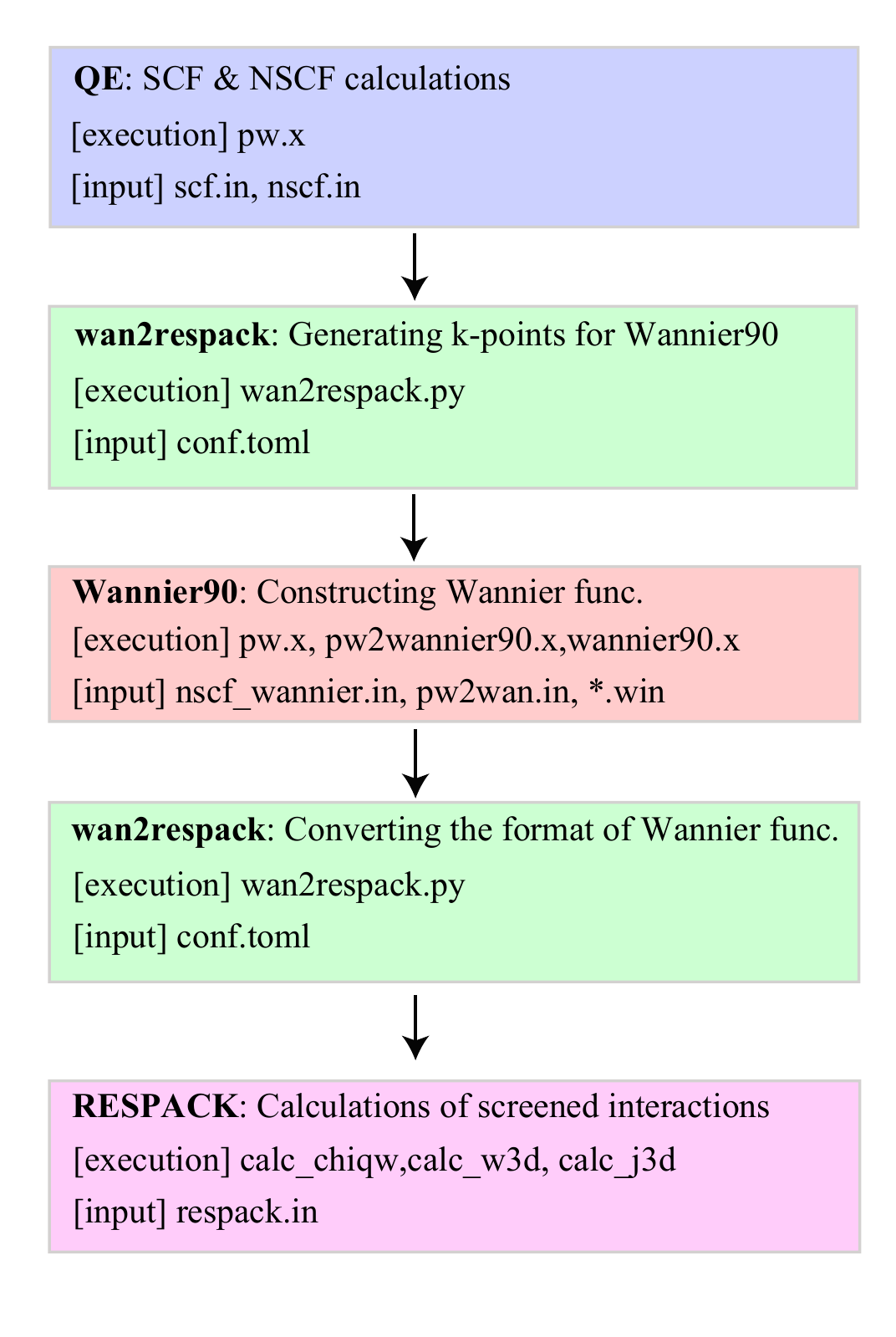}
    \caption{Schematic calculation flow. The necessary executions and input files in each step are also shown.
     }
    \label{fig:flow}
\end{figure}

The calculation flow of \software is as follows: 
\begin{enumerate}
\item $Ab$ $initio$ DFT calculations are performed using \verb|Quantum ESPRESSO|. 
\item ${\bm k}$ points are generated for \verb|Wannier90| using \verb|wan2respack|.
\item Wannier functions are constructed using \verb|Wannier90|.  
\item The Wannier functions are converted using \verb|wan2respack|. 
\item The screened interactions are calculated using \verb|RESPACK|. 
\end{enumerate}
Through 
these procedures, we obtain the screened Coulomb and exchange interactions in the low-energy effective Hamiltonians of solids.
This flow is summarized in Fig.~\ref{fig:flow}.

In the following, we explain the usage of \verb|wan2respack|. The input files for performing these calculations are in 
\verb|samples/seedname.lattice.kmesh|,
where \verb|seedname| is the name of the target compound, 
\verb|lattice| is the abbreviation of the 
Bravais lattice, and \verb|kmesh| is the number of $k$ points.
This directory consists of the following directories:
\begin{enumerate}
    \item \verb|PP|: This includes the pseudopotentials. 
    \item \verb|inputs|: This includes the input files for self-consistent field (scf) calculations (\verb|seedname.scf.in|), non self-consistent field (nscf) calculations (\verb|seedname.nscf.in|), Wannierization (\verb|seedname.pw2wan.in| and \verb|seedname.win.ref|), and the downfolding by \verb|RESPACK| (\verb|respack.in|). An input file for \verb|wan2respack| is also included (\verb|conf.toml|). The basic calculation flow is written in \verb|submit.sh|.
    \item \verb|inputs_selfk|: This includes the input files used by the user when setting ${\bm k}$ points for \verb|Wannier90|. The input files for \verb|Wannier90| (\verb|seedname.nscf_wannier.in| and \verb|seedname.win|) are prepared.
    \item \verb|reference|: This includes the reference input files for generating the Wannier functions using \verb|calc_wannier| in \verb|RESPACK| (\verb|RESPACK|-Wannier).
\end{enumerate}

\subsection{DFT calculations for irreducible k points}
DFT calculations of scf and nscf are performed using \verb|Quantum ESPRESSO| by executing the following commands:
\begin{verbatim}
$QE/bin/pw.x < seedname.scf.in 
             > seedname.scf.out
$QE/bin/pw.x < seedname.nscf.in 
             > seedname.nscf.out
\end{verbatim}
Here, \verb|$QE| indicates a path to the installation directory of \verb|Quantum ESPRESSO|. The calculated wave functions are used to compute the dielectric function using \verb|RESPACK|, so $\bm {k}$ points should be irreducible to reduce the computational cost.
\subsection{Export of k points to be calculated by Wannier90}
Next, for preprocessing, we generate a full $\bm k$-point list used by \verb|RESPACK|-Wannier. This ${\bm k}$-point list is exported to the input files of the nscf calculation and \verb|Wannier90| by executing the following commands:
\begin{verbatim}
$python $PATH_to_Install/bin/wan2respack.py 
-pp conf.toml
\end{verbatim}
The contents of \verb|conf.toml| are as follows:
\begin{verbatim}
[base]
QE_output_dir = "./work/seedname.save"
seedname = "seedname"
[pre.ref]
nscf = "seedname.nscf.in"
win = "seedname.win.ref"
[pre.output]
nscf = "seedname.nscf_wannier.in"
win = "seedname.win"
\end{verbatim}
\verb|conf.toml| should be written in the toml format~\cite{toml}. In the \verb|[base]| section, the output directory of \verb|Quantum ESPRESSO| and the seed name are specified using \verb|QE_output_dir| and \verb|seedname|, respectively. In the \verb|[pre.ref]| section, the reference files for generating the input files are specified using \verb|nscf| and \verb|win|, respectively. In the \verb|[pre.output]| section, the names of the output files generated by \verb|wan2respack| are indicated by \verb|nscf| and \verb|win|. After each calculation, the \verb|dir-wfn| directory and \verb|seedname.nscf_wannier.in| and \verb|seedname.win| files are generated (${\bm k}$ points are added to \verb|seedname.nscf_wannier.in| and \verb|seedname.win|).
\subsection{Generation of Wannier functions}
Wannier functions are generated using \verb|Quantum ESPRESSO|, \verb|Wannier90|, \verb|seedname.nscf_wannier.in|, and \verb|seedname.win| by executing the following commands:
\begin{verbatim}
$QE/bin/pw.x < seedname.nscf_wannier.in
             > seedname.nscf_wannier.out
$Wannier90/wannier90.x -pp seedname
$QE/bin/pw2wannier90.x < seedname.pw2wan.in 
                       > seedname.pw2wan.out
$Wannier90/wannier90.x seedname
\end{verbatim}
Here, \verb|$Wannier90| indicates a path to the installation directory of \verb|Wannier90|. 
\subsection{Conversion of Wannier functions into RESPACK format}
The Wannier functions obtained by \verb|Wannier90| are converted into the \verb|RESPACK| format by executing the following command:
\begin{verbatim}
$python $PATH_to_Install/bin/wan2respack.py 
conf.toml
\end{verbatim}
After these calculations, the following files are generated in the \verb|dir-wan| directory: 
\begin{itemize}
    \item \verb|dat.wan|: $\tilde{C}_{\bm G i}(\bm k)$ 
    \item \verb|dat.ns-nb|: $N_s, N_b$ 
    \item \verb|dat.umat|: $\sum_{m}U({\bm k})_{im} U^{\rm opt}({\bm k})_{mn}$ 
    \item \verb|dat.wan-center|: $\langle w_{i0}| {\bm r} | w_{i0}\rangle$ 
\end{itemize}

\subsection{Calculation of screened interactions using RESPACK}
The input file for \verb|RESPACK| is \verb|respack.in|. Using this file, we can calculate the screened Coulomb and exchange interactions through cRPA in \verb|RESPACK|. The execution command is as follows:
\begin{verbatim}
$RESPACK/bin/calc_chiqw < respack.in > LOG.chiqw
$RESPACK/bin/calc_w3d < respack.in > LOG.W3d
$RESPACK/bin/calc_j3d < respack.in > LOG.J3d
\end{verbatim}
For comparison, we provide an input file for \verb|RESPACK|-Wannier in \verb|reference/respack.in|. As shown later, the obtained Wannier functions and screened interactions by \verb|Wannier90| and \verb|RESPACK| quantitatively agree.

\begin{figure}
    \centering
    \includegraphics[width=\columnwidth]{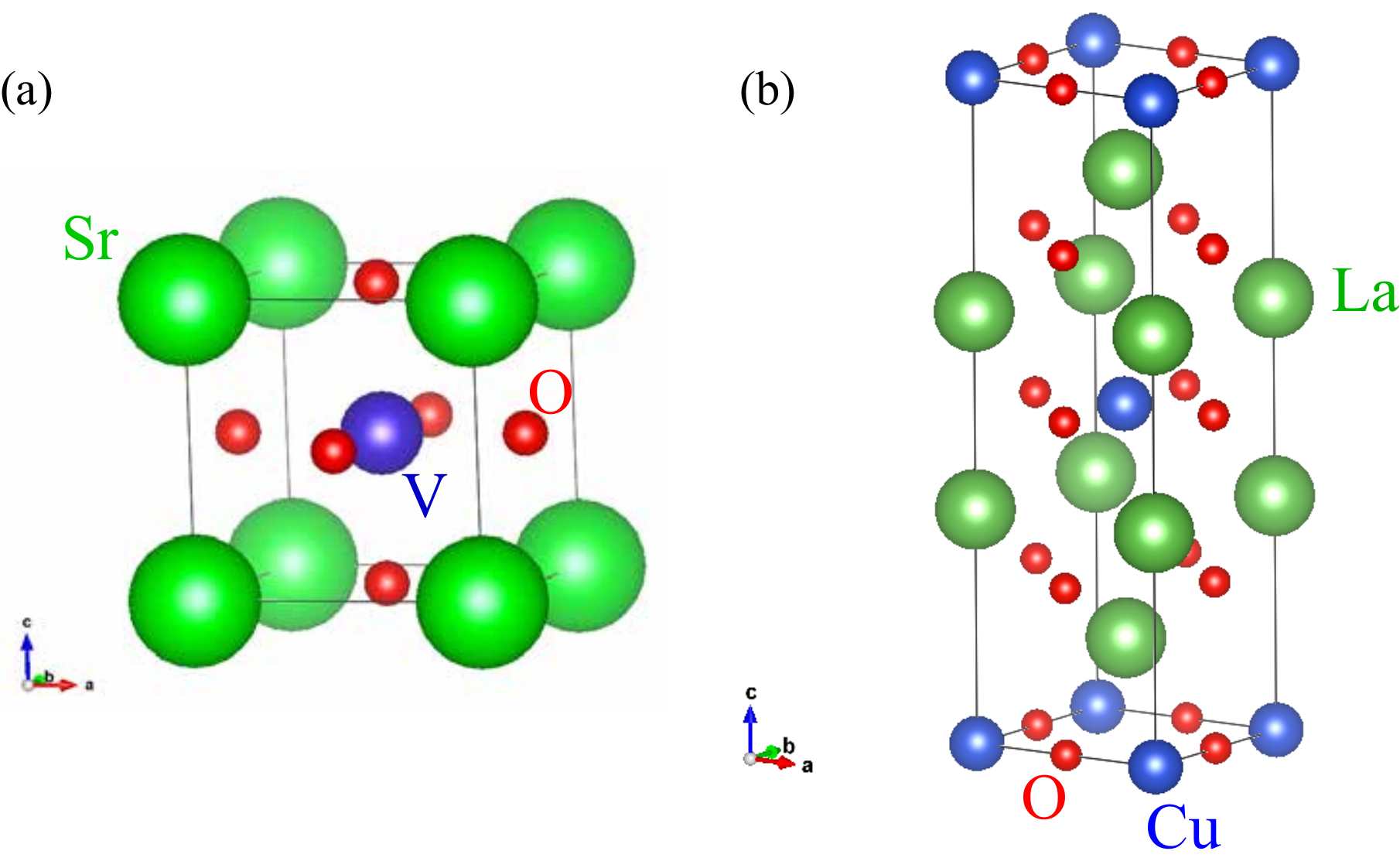}
    \cprotect\caption{Crystal structures of (a) SrVO$_3$  and (b) La$_2$CuO$_4$ visualized by using {\tt VESTA}\cite{vesta}. The space groups of SrVO$_3$ and La$_2$CuO$_4$ shown here are Pm$\bar{3}$m and I4/mmm, respectively.
}
    \label{fig:structure}
\end{figure}

\begin{figure}[t]
    \centering
    \includegraphics[width=\columnwidth]{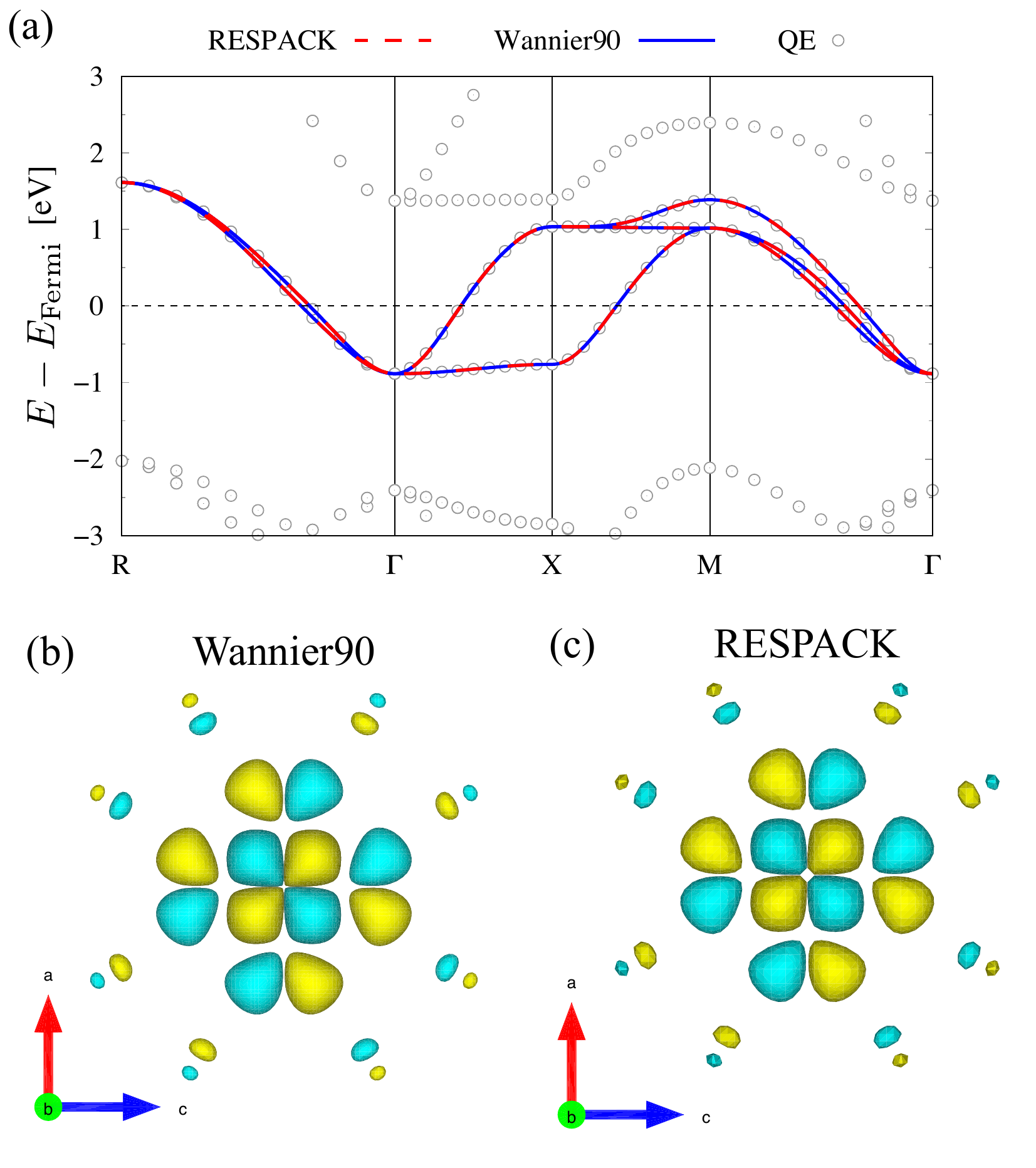}
    \cprotect\caption{(a)~Band structure of SrVO$_3$ interpolated by Wannier functions. The blue and red lines 
show the results obtained using \verb|Wannier90| and \verb|RESPACK|, respectively. 
$E_{\rm Fermi}$ denotes the Fermi energy. The dashed black line represents the Fermi level. 
The coordinates of the high-symmetry $\bm{k}$-points in the horizontal axis are ${\rm \Gamma} = (0, 0, 0)$, R=(0.5, 0.5, 0.5), X=(0.5, 0, 0) and M=(0.5, 0.5, 0). 
The gray circles depict the band structure obtained using \verb|Quantum ESPRESSO|.
Wannier functions with $\alpha=1$ of SrVO$_3$ obtained by (b) \verb|Wannier90| and (c) \verb|RESPACK|. 
The square root of the volume $1/\sqrt{\Omega}$ is included in the definition of Wannier 
functions in \verb|RESPACK| but not in the definition of Wannier functions in \verb|Wannier90|. 
See also Eq. (\ref{eq:wan_respack}).
This figure is visualized using \verb|VESTA|\cite{vesta}.
}
    \label{fig:SVO-band}
\end{figure}

\section{Application Examples}
In this section, we show the application of \verb|wan2respack| to typical compounds for correlated electron systems, namely, SrVO$_3$ and La$_2$CuO$_4$, whose crystal structures are shown in Fig. \ref{fig:structure}. The inputs are uploaded to the GitHub repository as samples. Some outputs in this section are also available elsewhere\cite{issp_gitlab}. Note that we use only the disentanglement scheme when we construct the Wannier functions by \verb|Wannier90|. 
Nevertheless, as we show later, the Wannier functions constructed from \verb|Wannier90| and \verb|RESPACK| show good agreement with each other.

\subsection{SrVO$_3$}
SrVO$_3$ is a typical correlated metal compound\cite{onoda1991metallic,Inoue_PRB1998}. Three $t_{2g}$-like orbitals are located around the Fermi energy and isolated from the other bands\cite{Lechermann_PRB2006}. As seen in Fig. \ref{fig:SVO-band}(a), the Wannier bands of the $t_{2g}$ orbitals obtained using \verb|RESPACK| and \verb|Wannier90| reproduce the original DFT band structure well. Figure \ref{fig:SVO-band}(b) and (c) show Wannier functions with an orbital index $\alpha=1$. 
The sum of the spreads of all the Wannier orbitals are 5.5161 $[{\rm \AA^2}]$ for  \verb|Wannier90|
and  5.5156 ${[\rm \AA^2}]$ for \verb|RESPACK|\cprotect\footnote{As the unit of the spread, the square of the Bohr radius $a_0^2$ is employed in the outputs of \verb|RESPACK|.}. 
These results suggest that both software can generate the same Wannier orbitals quantitatively. 

\begin{figure}
    \centering
    \includegraphics[width=\columnwidth]{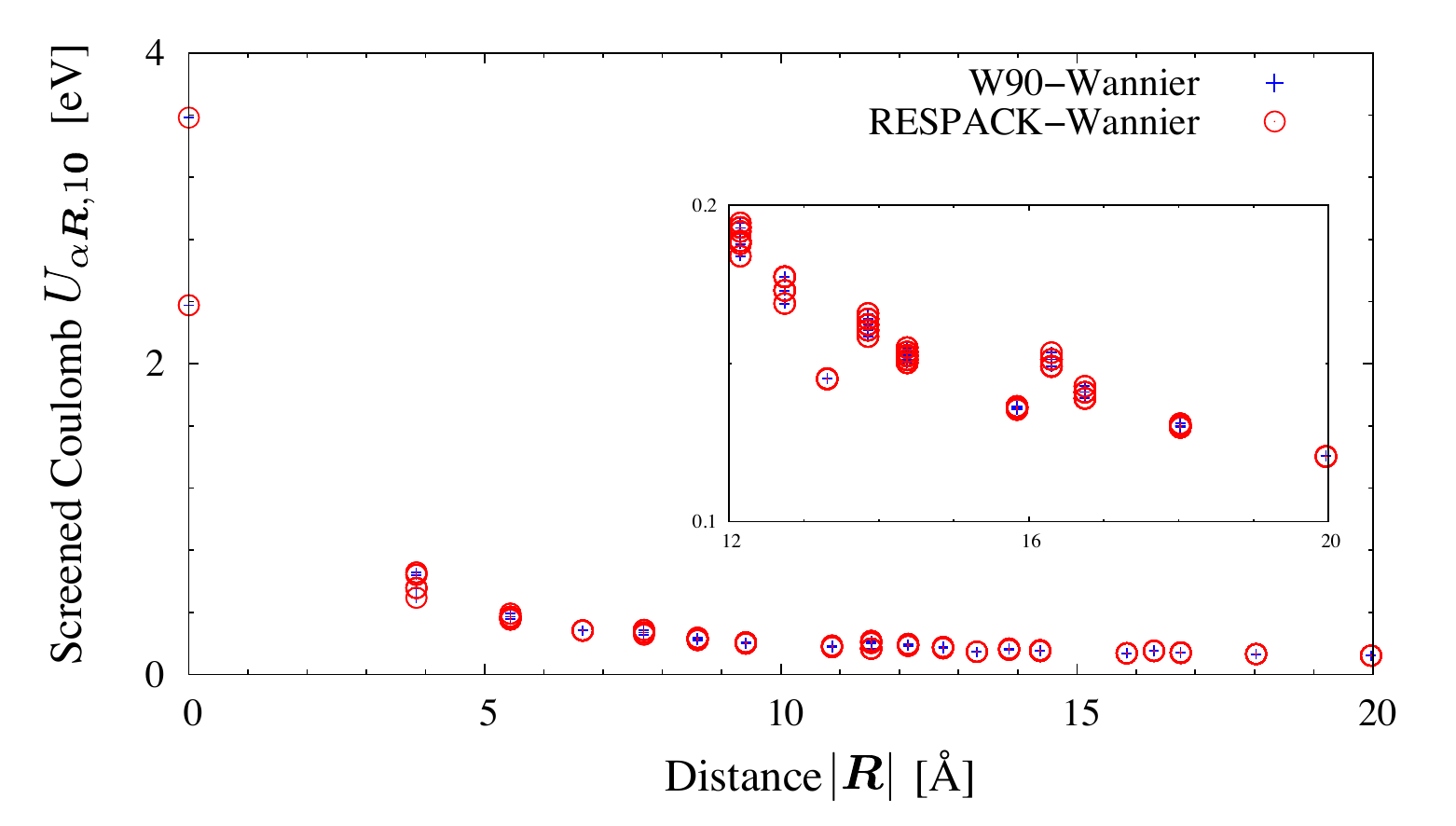}
    \cprotect\caption{Distance dependence of screened Coulomb interactions $U_{\alpha\vec{R},\beta \vec{R}'}$ obtained via cRPA for $\beta=1$ and $\vec{R}'=\vec{0}$. W90-Wannier and RESPACK-Wannier indicate the numerical findings from using the Wannier functions obtained by \verb|Wannier90| and \verb|RESPACK|. The inset shows an enlarged view of the Coulomb interactions.
}
    \label{fig:SVO-cRPA}
\end{figure}

Since \verb|RESPACK| and \verb|Wannier90| obtain almost the same Wannier functions, the effective interactions obtained using either Wannier function should be the same if \verb|wan2respack| works correctly. To confirm this statement, we plot an effective Coulomb interaction written in \verb|dat.WvsR.001| in Fig. \ref{fig:SVO-cRPA}. The two results show excellent agreement, which indicates that \verb|wan2respack| works well.

\subsection{La$_2$CuO$_4$}
We show the application of \verb|wan2respack| to La$_2$CuO$_4$, which is a 
parent compound of high-$T_{\rm c}$ cuprate superconductors\cite{bednorz1986possible, keimer2015quantum}. Although this compound is a Mott insulator of an antiferromagnetic order, most of the DFT results suggest that this compound is a paramagnetic metal in its ground state\cite{Mattheiss_PRL1987, Pickett_RMP1989}. 
In contrast to the DFT calculations, 
it is shown that the Mott insulating state can be reproduced 
by solving the effective Hamiltonians derived 
from $ab$ $initio$ downfolding~\cite{Hirayama_PRB2019, Ohgoe_PRB2020}. 

\begin{figure}[t]
    \centering
    \includegraphics[width=\columnwidth]{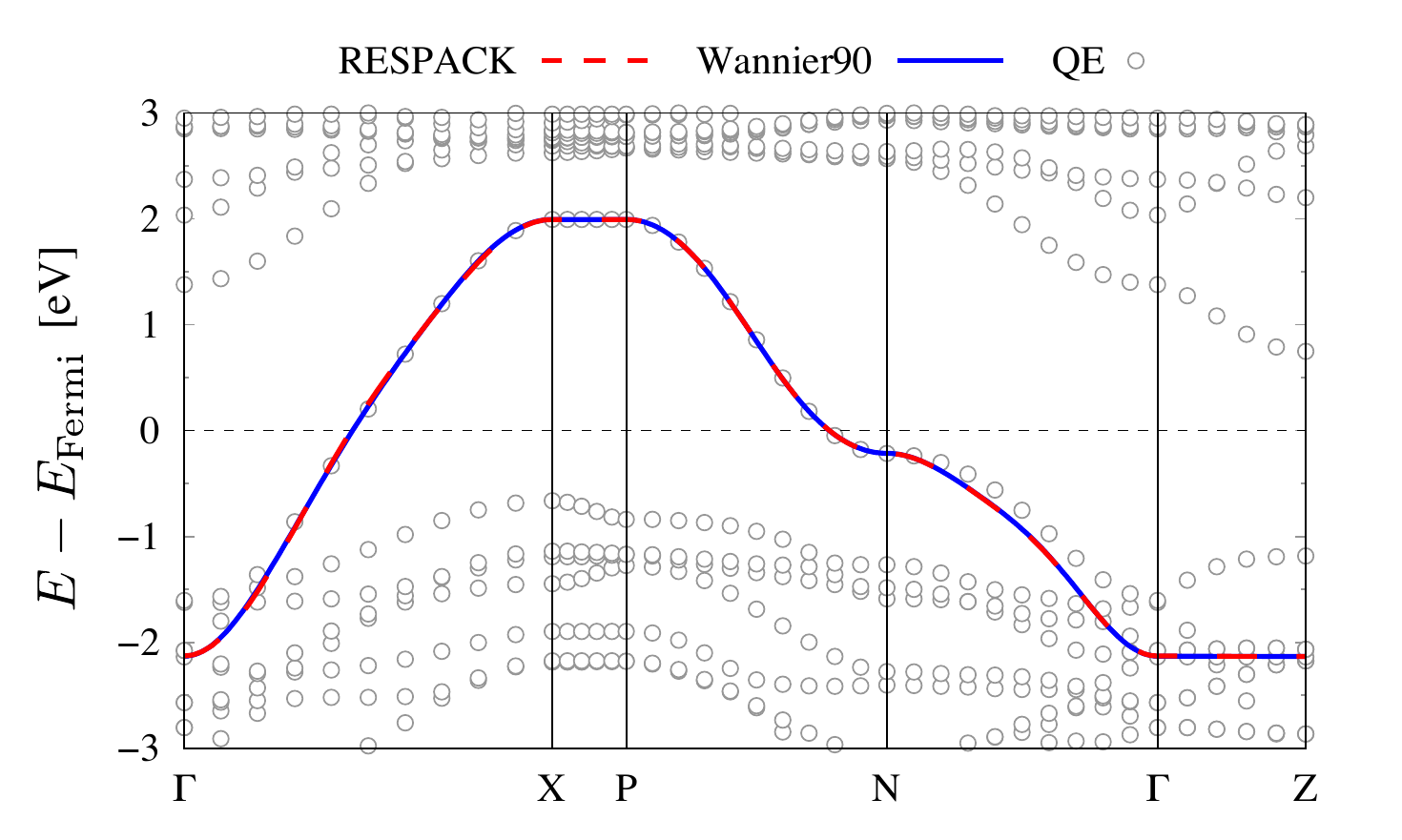}
    \caption{Band structure of La$_2$CuO$_4$ interpolated using Wannier functions. The coordinates of the high-symmetry $\bm{k}$-points in the horizontal axis are ${\rm \Gamma} = (0, 0, 0)$, X=(0, 0, 0.5), P=(0.25, 0.25, 0.25), N=(0, 0.5, 0), and Z=(0.5, 0.5, -0.5). The notations are the same as in Fig. \ref{fig:SVO-band}(a).}
    \label{fig:La2CuO4-band}
\end{figure}

\begin{figure}[t]
    \centering
    \includegraphics[width=\columnwidth]{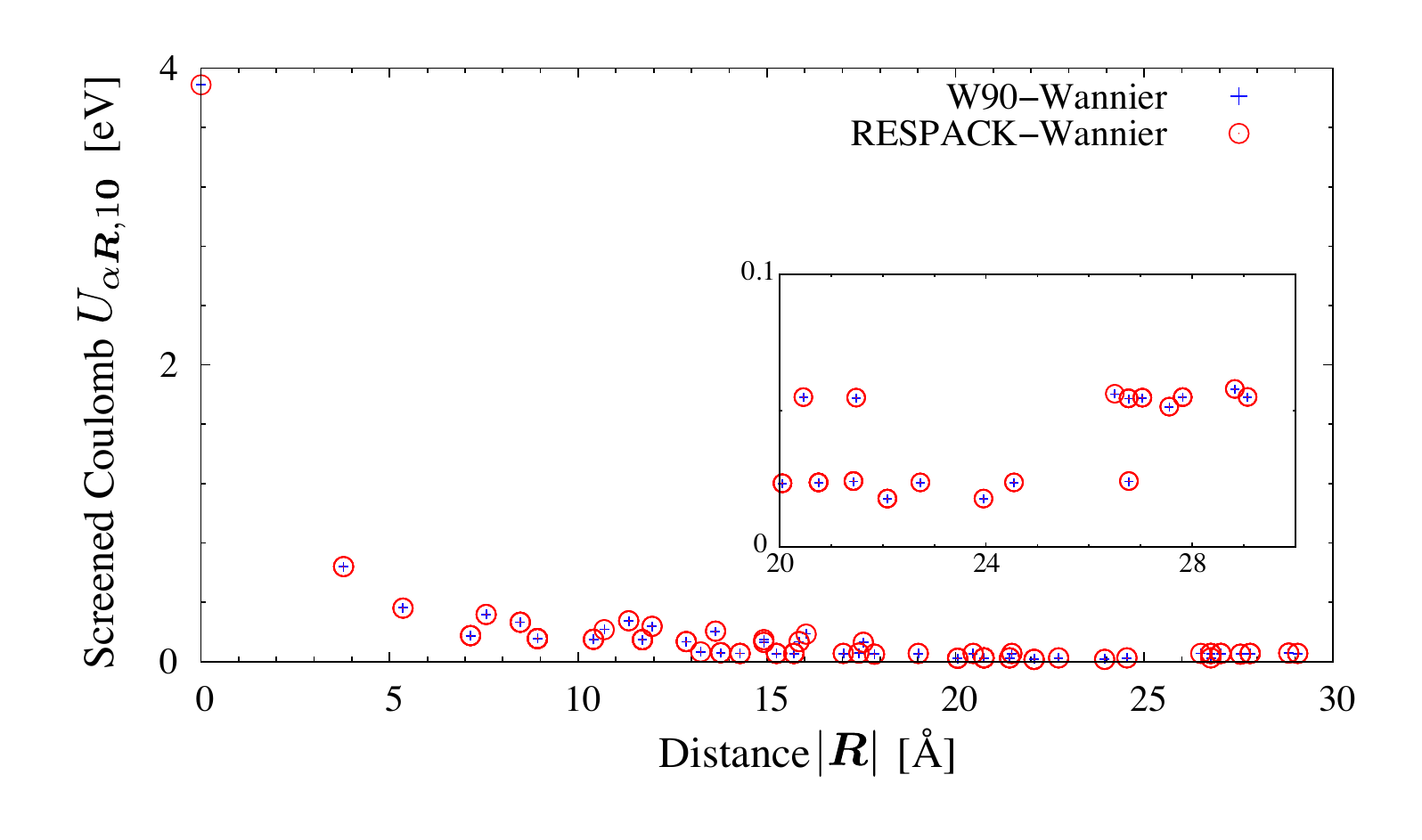}
    \caption{Distance dependence of screened Coulomb interactions $U_{\alpha\vec{R},1\vec{0}}$ for La$_2$CuO$_4$ obtained via cRPA. We plot only the data for $|\bm{R}|\leq 30$\cite{note_cRPA}. The notations are the same as in Fig. \ref{fig:SVO-cRPA}. }
    \label{fig:La2CuO4-cRPA}
\end{figure}

Compared with SrVO$_3$, La$_2$CuO$_4$ has a more complicated band structure\cite{Mattheiss_PRL1987, Pickett_RMP1989, Hirayama_PRB2019}. The $d_{x^2-y^2}$ orbital of copper hybridized with the $p$ orbitals of oxygen crosses the Fermi energy. Contrary to that of SrVO$_3$, this band is not well isolated: the other bands, such as the $d_{3z^2-r^2}$ orbital of copper and the $p$ orbitals of oxygen, exist near the Fermi energy. Such complexity of band structures sometimes leads to difficulties in constructing Wannier orbitals. 
Nevertheless, as shown in Fig. \ref{fig:La2CuO4-band}, 
the band structures near the Fermi level generated 
by \verb|RESPACK| and \verb|Wannier90| are consistent with each other. 
We also confirm that the Wannier spreads are also consistent with each other 
(4.0231 [${\rm \AA^2}$] for \verb|Wannire90| and 4.0235  [${\rm \AA^2}$] for \verb|RESPACK|).
Moreover, we confirm that the effective interactions calculated via \verb|wan2respack| 
agree well with the original \verb|RESPACK| results, as shown in Fig. \ref{fig:La2CuO4-cRPA}.

\section{Summary}
We developed the interface tool \verb|wan2respack|, which converts Wannier functions in the \verb|Wannier90| format into those in the \verb|RESPACK| format. Using \verb|wan2respack|, one can perform \verb|RESPACK| calculations using the Wannier functions obtained by \verb|Wannier90|. For example, one can obtain the $ab$ $initio$ low-energy effective Hamiltonians of solids using \verb|RESPACK|. To demonstrate the use of \verb|wan2respack|, we derive low-energy effective Hamiltonians for the correlated metal SrVO$_{3}$ and the high-$T_{c}$ superconductor La$_{2}$CuO$_{4}$ using Wannier functions obtained by \verb|Wannier90| and those obtained by \verb|RESPACK|-Wannier. From these applications, we confirm that the low-energy effective Hamiltonians derived through both implementations of Wannier functions are the same. Since \verb|Wannier90| is a standard tool for constructing Wannier functions, the connection between \verb|Wannier90| and \verb|RESPACK| via \verb|wan2respack| further enhances the usability of \verb|RESPACK|.

\section*{Acknowledgments}
We are indebted to Kazuma Nakamura for valuable discussions and for providing us with several codes.
We acknowledge Tetsuya Shoji, Noritsugu Sakuma, and Tetsuya Fukushima for fruitful discussions and important suggestions. A part of this work is financially supported by TOYOTA MOTOR CORPORATION. KY and TM were supported by Building of Consortia for the Development of Human Resources in Science and Technology from the MEXT of Japan. 
We thank Oakbridge-CX in the Information Technology Center, the University of Tokyo and the Supercomputer Center, the Institute for Solid State Physics, the University of Tokyo for their facilities.
This work has also been supported by JST-Mirai Program (JPMJMI20A1), Grant-in-Aid for Scientific Research (Nos. 21H01003, 21H01041, 21H04437, 22K03447, and 22K18954) from Ministry of Education, Culture, Sports, Science and Technology, Japan, and
the National Natural Science Foundation of China (Grant No. 12150610462).

\bibliographystyle{elsarticle-num}
\bibliography{main}

\begin{thebibliography}{10}
\bibitem{note_cRPA}
The long-range part of the screened Coulomb interactions for $|\bm{R}| > 30$
  may be meaningless due to the boundary effect of the supercell.
\end{thebibliography}

\end{document}